\documentclass[11pt,titlepage,a4paper]{article}
\usepackage{bozhomac}	
\usepackage{bozhlogo}	
\usepackage{cite}	

\title{\bfseries	\vspace*{-1.678902345in}
{\huge   Deviation equations and\\[1ex] weak equivalence principle\\[1ex]
	 in spaces with affine connection        }
}

\vspace{1.7ex}

\author{
Bozhidar Z.\ Iliev
\thanks{Laboratory of Mathematical Modeling in Physics,
Institute for Nuclear Research and \mbox{Nuclear} Energy,
Bulgarian Academy of Sciences,
Boul.\ Tzarigradsko chauss\'ee~72, 1784 Sofia, Bulgaria}
\thanks{E-mail address: bozho@inrne.bas.bg}
\thanks{URL: http://theo.inrne.bas.bg/$\sim$bozho/}
}

\date{
 \vspace{2.27ex}\ShortTitle{Deviation equations and equivalence principle}	\\[0.27ex]
 \vspace{3.27ex}
\small
	\begin{tabular}{r@{$\colon\to~$}l}
 \vspace{0.27ex} Produced	& \fbox{\today}	\\[0.27ex]
	\end{tabular} \\[1.27ex]
\normalsize
	\begin{tabular}{r@{$\colon~$}l}
 \vspace{0.27ex} http://arXiv.org e-Print archive No. & gr-qc/0605008
								\\[0.27ex]
\vspace{0.27ex} Published in	       
Proceedings  of the 6-th Soviet (USSR) Gra\-vity Conference\\
``Modern  theoretical and experimental problems of relativity theory\\
and gravitation'', Univ. "Friendship between nations"\\
Moscow, 3-5 July 1984,  Moscow, 1984, p.~227 (In Russian).	\\[0.27ex]
	\end{tabular} \\[-0.27ex]
 \vspace{4.27ex}{\Huge\BOZHO}	\\[4.27ex]
 \vspace{0.27ex}\Subject{General relativity}	\\[2.27ex]
	\begin{tabular}{r@{\hspace{0.512em}}|@{\hspace{0.512em}}l}
 \vspace{0.27ex}	
&
 \vspace{0.27ex} 
	\end{tabular} \\[1.27ex]
\vspace{0.27ex}	\\[0.27ex]
}

\begin{document}		

\renewcommand{\thepage}{\roman{page}}

\renewcommand{\thefootnote}{\fnsymbol{footnote}} 
\maketitle				
\renewcommand{\thefootnote}{\arabic{footnote}}   




\renewcommand{\thepage}{\arabic{page}}



\renewcommand{\theequation}{\arabic{equation}}

\vspace*{3ex}
	\begin{center}
	\begin{minipage}{0.6\textwidth}
\textbf{Abstract.}
Some connections between the deviation equations and weak equivalence
principle are investigated.\\[5ex]
	\end{minipage}
	\end{center}

	For any point $x_0$ in a space with affine connection $L_n$, there
exists generally anholonomic frame $(i_0)$ such that the connection
coefficients $\Gamma_{i_0j_0}^{k_0}$ vanish at $x_0$ in it,
$\Gamma_{i_0j_0}^{k_0}(x_0)=0$, but, in the general case, their derivatives
do not vanish, $\Gamma_{i_0j_0,l_0}^{k_0}(x_0)\not=0$
(see~\cite{bp-DE+Sava,Heyde}).  This system/frame is  called locally inertial
or locally Lorentzian.

	The weak equivalence principle states
that~\cite{MTW,Mitskevich,Heyde} in the above pointed frame in
``sufficiently small'' neighborhood of the point $x_0$ the laws/equations of
motion take a form identical (up to infinitesimal quantities) with the one in
(locally-)flat spacetime. Note, this can be exactly valid at the point $x_0$
and possibly along some path through it.

	It can be asserted that the (generalized) deviation
equation~\cite{bp-DE+Sava} is the rigorous mathematical equivalent to the
weak equivalence principle. Moreover, that assertion is valid on arbitrary
spaces $L_n$ with affine connection.

	To illustrate the above statement, consider the deviation equation of
two free, ``infinitesimally close'', and independent
particles~\cite{bp-DE+Sava}:
	\begin{equation}	\label{1}
\frac{\bar{D}V^k}{\od\tau}
=
R_{ijl}^{k} u^iu^j\xi^l + u^j \frac{\bar{D} (T_{jl}^{k}\xi^l)}{\od\tau} ,
	\end{equation}
where $\tau$ is the parameter of the trajectory (worldline) of one of the
particles (basic particle or observer) with tangent vector $u^i$,
$\frac{\bar{D}}{\od \tau}$ is the covariant derivative with respect to
$\tau$, $R_{ijl}^{k}$ is the curvature tensor,
\(
T_{jl}^{k}
:=  - 2 \Gamma_{[il]}^{k} - C_{il}^{k}
\)
with $C_{il}^{k}$ defining the commutators of the basic vectors of the
frames and $[\ldots]$ denoting antisymmetrization with coefficient
$\frac{1}{2}$, $\xi^i$ is the (infinitesimal) displacement vector of the
particles, and $V^k:=\frac{\bar{D}\xi^k}{\od \tau}$ is their relative
velocity.

	In (locally) flat  $L_n$ space, the equation~\eref{1} takes the form
	\begin{equation}	\label{2}
\frac{\bar{D}V^k}{\od\tau}
= u^j \frac{\bar{D}(T_{jl}^{k}\xi^l)}{\od\tau}
= T_{jl}^{k}  u^j  V^l
+ u^j  \xi^l\frac{\bar{D} T_{jl}^{k}}{\od\tau}
	\end{equation}
at every spacetime point and in any frame $(i)$.
The equations~\eref{1} and~\eref{2} in a locally Lorentzian frame $(i_0)$ at
$x_0$ read respectively
	\begin{gather}	\label{3}
\frac{ \bar{D}V^{k_0} }{\od\tau} \Big|_{x=x_0}^{(1)}
=
\Bigl[ \Bigl(
- 2 \Gamma_{i_0 (j_0,l_0)}^{k_0}  u^{i_0}  \xi^{l_0}
- C_{j_0l_0}^{k_0} V^{l_0}
+ \xi^{l_0} \frac{\bar{D} T_{j_0l_0}^{k_0}}{\od\tau}
\Bigr) u^{j_0} \Bigr] \Big|_{x=x_0}
\\			\label{4}
\frac{ \bar{D}V^{k_0} }{\od\tau} \Big|_{x=x_0}^{(2)}
=
\Bigl[ \Bigl(
- C_{j_0l_0}^{k_0} V^{l_0}
+ \xi^{l_0} \frac{\bar{D} T_{j_0l_0}^{k_0}}{\od\tau}
\Bigr) u^{j_0} \Bigr] \Big|_{x=x_0}
	\end{gather}

	Let us introduce the quantities
	\begin{equation}
A_{k_0}(x_0)
:=
\frac{ \bar{D}V^{k_0} }{\od\tau} \Big|_{x=x_0}^{(1)}
-
\frac{ \bar{D}V^{k_0} }{\od\tau} \Big|_{x=x_0}^{(2)}
=
[(\cdots)u^{j_0} \xi^{l_0}]|_{x=x_0} ,
	\end{equation}
which are generally not components of a vector.

	The weak equivalence principle states in the particular case that in
``sufficiently small'' neighborhood of $x_0$ (which is assumed to be on the
basic path) the r.h.s.\ of~\eref{3} and~\eref{4} must coincide, \ie
$A^{k_0}(x_0)=0$. Excluding some particular cases (like $\xi^i=\const u^i$),
this means that the particles coincide at the point $x_0$, \ie
$\xi^l_0(x_0)=0$. This result agrees with the fact that the weak equivalence
principle is a local statement. Therefore the quantities $A^{k_0}(x_0)$ are a
measure for the validity of the weak equivalence principle in a neighborhood
of any spacetime point.

	It is clear in the particular case, the weak equivalence principle
is completely contained in the deviation equation. Similar consideration
reveal that the weak equivalence principle is a consequence of the
generalized deviation equation (cf.~\cite{bp-DE+Sava}).

\addcontentsline{toc}{section}{References}
\bibliography{bozhopub,bozhoref}
\bibliographystyle{unsrt}
\addcontentsline{toc}{subsubsection}{This article ends at page}

\end{document}